\documentclass[aps,11pt,final,
notitlepage, oneside,  nobibnotes, nofootinbib,
superscriptaddress, showpacs, centertags, showkeys, assume]{revtex4}

\usepackage{epsfig}
\usepackage{graphicx}
\usepackage{amsfonts}
\usepackage{amsmath}
\usepackage{epstopdf}
\begin{document}

\title{Pion scalar form factor and model independent values of $f_0(500)$ and $f_0(980)$ meson parameters}

\author{S.Dubni\v cka}
\address{Institute of Physics, Slovak Academy of Sciences,
Bratislava, Slovak Republic}
\author{A.Z.Dubni\v ckov\'a}
\address{Dept. of Theoretical Physics, Comenius University, Bratislava,
Slovak Republic}
\author{A. Liptaj}
\address{Institute of Physics, Slovak Academy of Sciences,
Bratislava, Slovak Republic}


\vspace{2cm}
\begin{abstract}
  The pion scalar form factor along with all existing S-wave iso-scalar
$\pi\pi$-scattering phase shift data in the elastic region up to
$1 {\rm GeV}^2$ are applied for a determination of the $f_0(500)$ and
$f_0(980)$ meson parameters $m_{f_0(500)}=(360\pm33){\rm MeV}$,
$\Gamma_{f_0(500)}=(587\pm 85){\rm MeV}$ and
$m_{f_0(980)}=(957\pm77) {\rm MeV}$, $\Gamma_{f_0(980)}=(164\pm142){\rm MeV}$
in a model independent way.
\end{abstract}

\maketitle

\section{INTRODUCTION}

   In contrast to other $SU(3)$ multiplets of hadrons, the
identification of the scalar mesons is a long-standing puzzle as
some of them have decay widths which cause a strong overlap
between resonances and background. The situation below 1 GeV is
even more complicated because scalar mesons possess identical
quantum numbers with glueballs $0^{++}$ which can appear just in
this mass region.

   Despite of this fact, all experimentally established scalar mesons
\cite{beringer} are now classified  into light scalar nonet
comprising the $f_0(500), K_0^*(800), f_0(980)$ and $a_0(980)$
mesons, not necessarily to be $q \bar q$ states, and into regular
nonet consisting of the $f_0(1370), K_0^*(1430), a_0(1450)$ and
$f_0(1500)$ (or $f_0(1700)$) mesons.

   It has been difficult to establish the precise $f_0(500)$ meson
parameters because of its large width and because it can certainly
not to be determined by a naive Breit-Wigner form. As a result
historically this so-called $\sigma$ meson has been listed in PDG
as "not well established" until 1974 and one has believed in the
existence of a broad and light scalar iso-scalar resonance.

   However, it has been removed from PDG in 1976, because two heavier
resonances $f_0(980)$ and $f_0(1300)$ were found and $\sigma$
meson could be replaced by these two heavier ones, which could
complete a light $(q \bar q)$ nonet.

   Then again listed back in 1996, after missing more than two
decades, although still with an obscure denotation
$f_0(400-1200)$.

   From 2002 as "well established" $f_0(600)$, but with
conservative estimate of the mass $400-1200 {\rm MeV}$ and the width
$600-1000 {\rm MeV}$.

   A clarification of this controversial situation has been
achieved only recently in the paper \cite{gar-mar}, where also
other recent determinations of reasonable $f_0(500)$ parameters
are presented.

   In this paper we confirm the existence of the $f_0(500)$
scalar meson in a model independent way. As for the latter a
representation of the pion scalar form factor is valid in the
whole elastic region up to $1 {{\rm GeV}}^2$, one can determine also the
$f_0(980)$ meson parameters as well.

  With this aim, by means of the unitary and analytic approach,
an explicit form of the pion scalar form factor (FF) in the
language of the absolute value of the pion c.m. three-momentum
variable $q$, to be connected with the momentum transfer squared
by the relation $t=4(q^2+m^2_\pi)$, is constructed. As a result
the pion scalar FF takes the form of a rational function in
$q$-plane and some of its poles give masses and widths of
$f_0(500)$ and $f_0(980)$ resonances.

 There is a general belief that the $f_0(500)$ and $f_0(980)$ mesons are the
lowest members of the light scalar nonet, however, it is not
definitely clear what other particles are really members of it.

\section{THE PION SCALAR FORM FACTOR}

   The pion scalar FF $\Gamma_\pi(t)$ is defined by the matrix element of the
quark density
\begin{equation}
  <\pi^i(p_2)\mid \widehat{m} (\bar u u + \bar d d)\mid \pi^j(p_1)> =
  \delta^{ij}\Gamma_\pi(t)
\end{equation}
where $t=(p_2-p_1)^2$, $\widehat{m}=\frac{1}{2}(m_u + m_d)$ and it
has similar properties to the pion electromagnetic FF
\cite{dubn1}.

   The pion scalar FF $\Gamma_{\pi}(t)$ is analytic in the whole
complex $t$-plane, except for a cut along the positive real axis,
starting at $t=4 m_{\pi}^2$.

   For real values $t< 4 m_{\pi}^2$  $\Gamma_{\pi}(t)$ is real. The
latter implies the so-called reality condition $\Gamma^*_{\pi}(t)$
= $\Gamma_{\pi}(t^*)$, i.e. that the values of FF above and below
the cut are complex conjugate of each other.

   At $t=0$ the pion scalar FF $\Gamma_{\pi}(t)$ coincides with
the pion sigma-term \cite{gasser} $\Gamma(0)=(0.99\pm0.02)m_\pi^2$
to be evaluated in the framework of chiral perturbation theory.
Further, the pion scalar FF will be normalized exactly to
$m_\pi^2$.

   The FF $\Gamma_{\pi}(t)$ is not directly measurable quantity and it
enters e.g. in the matrix element for the decay of the Higgs boson
into two pions. However, the contribution to the decay rate seems
to be negligible small \cite{shifman,voloshin}.

   If $\Gamma_{\pi}(t)$ is evaluated on the upper boundary of the
cut, one finds that the following unitarity condition is obeyed

\begin{equation}
  Im \Gamma_{\pi}(t) = \sum_n
  <\pi(p')\pi(p)\mid T \mid n><n\mid \widehat{m}(\bar u u + \bar d
  d)\mid 0>\label{unitarity}
\end{equation}
where $T$ is the $T$-operator and the sum runs over a complete set
of allowed states $2\pi, 4\pi,...,K \bar K, ...$, which create
additional branch points on the positive real axis of the
$t$-plane between $4m_\pi^2$ and $\infty$.

   In the elastic region $4m_\pi^2\leq t \leq 16m_\pi^2$ only the
first term on the right hand side of (\ref{unitarity}) contributes
and then
\begin{equation}
  Im \Gamma_{\pi}(t) = \Gamma_{\pi}(t)(\sigma T^0_0)^*  \label{elasticunit1}
\end{equation}
where $\sigma T^0_0$ is the $S$-wave iso-scalar $\pi \pi$
scattering amplitude
\begin{equation}
  M^0_0 = \sigma T^0_0 = \frac{1}{2i}(e^{2i \delta} -1); \label{ampl}
\end{equation}
$\delta = \delta^0_0 +i \varphi;$ $\delta^0_0, \varphi$ real,
where $\delta^0_0$ stands for the $S$-wave iso-scalar $\pi \pi$
phase shift, and $\varphi>0$ measures the inelasticity.

   Though $\varphi$ exactly vanishes only below $t = 16m_\pi^2$,
indeed the phenomenological analysis of the $\pi \pi$ interactions
\cite{mmsh} shows that final states containing more than two
particles start playing a significant role only well above $
4m^2_K \approx 1 {\rm GeV}^2$, where the inelastic two-body channel $\pi
\pi \to K \bar K$ opens.

   Then
\begin{equation}
   Im \Gamma_{\pi}(t) =
   \Gamma_{\pi}(t)e^{-i\delta_0^0}sin \delta_0^0 \label{elasticunit2}
\end{equation}
for $4m^2_\pi \leq t \leq 1 {\rm GeV}^2$.

   From the relation (\ref{elasticunit2}) it follows that the phase
$\delta_\Gamma$ of $\Gamma_{\pi}(t)$ coincides with $\delta^0_0$
and just this identity enables us to obtain the pion scalar FF
$\Gamma_{\pi}(t)$ behavior valid at the elastic interval $4m^2_\pi
\leq t \leq 1 {\rm GeV}^2$ and subsequently allows the identification of
the $f_0(500)$ meson pole and the pole of $f_0(980)$ meson as well
on the second Riemann sheet in $t$-variable.

   The asymptotic behavior
\begin{equation}
   \Gamma_{\pi}(t)_{\mid t \mid \to \infty} \sim \frac{1}{t}
\end{equation}
is predicted by the quark counting rules.

   Starting from the unitarity condition for the $S$-wave
iso-scalar $\pi\pi$ scattering amplitude
\begin{equation}
   Im M^0_0 =-|M^0_0|^2
\end{equation}
one can do analytic continuation of $M^0_0$ through the upper and
lower boundaries of the unitary elastic cut and to prove in this
way  the singularity at $t = 4m_\pi^2$ to be a square root branch
point. As a result one gets
\begin{equation}
   (M^0_0)^{II} = \frac{(M^0_0)^I}{1-2i (M^0_0)^I}\label{amplsecsheet}.
\end{equation}

   The same can be done with the pion scalar FF and as a result
one gets the expression
\begin{equation}
   (\Gamma_\pi)^{II} = \frac{(\Gamma_\pi)^I}{1-2i
   (M^0_0)^I}\label{ffsecsheet},
\end{equation}
relating the pion scalar FF on the second Riemann sheet with the
pion scalar FF and the $S$-wave iso-scalar ${\pi \pi}$ scattering
amplitude on the first Riemann sheet, demonstrating in this way
that the singular point of pion scalar FF at $t = 4m_\pi^2$ is
square root branch point, generating two sheeted Riemann surface
on which the pion scalar FF is defined.

   Moreover, by a comparison of (\ref{amplsecsheet}) with
(\ref{ffsecsheet}) one can see that both expressions have
identical denominator, from where it automaticly follows that if
there are $f_0(500)$ and $f_0(980)$ mesons in the form of the
poles of the $S$-wave iso-scalar $\pi\pi$ scattering amplitude on
the second Riemann sheet, then they appear also as poles on the
second Riemann sheet of the pion scalar FF.

   Now, by an application of the conformal mapping
\begin{equation}
   q = [(t-4)/4]^{1/2},\quad m_\pi = 1
\end{equation}
two-sheeted Riemann surface of $\Gamma_\pi(t)$ is mapped into one
absolute valued pion c.m. three-momentum $q$-plane and the elastic
cut disappears. Neglecting all higher branch points, there are
only poles and zeros of $\Gamma_\pi(t)$ in $q$-plane and as a
consequence the pion scalar FF can be represented by a Pad`e-type
approximation
\begin{equation}
   \Gamma_\pi(t) = \frac{\sum_{n=0}^M a_n
   q^n}{\prod_{i=1}^N(q-q_i)}\label{padet}.
\end{equation}
Because $\Gamma_\pi(t)$ is a real analytic function, the
coefficients $a_n$ in (\ref{padet}) with $M$ even (odd) are real
(pure imaginary), respectively, and the poles $q_i$ can appear on
the imaginary axis or they are placed always two of them
symmetrically according to it.

   If one multiplies both, the numerator and the denominator of
(\ref{padet}), by the complex conjugate factor
$\prod_{i=1}^N(q-q_i)^*$, the new denominator is a polynomial with
real coefficients already and tangent of the pion scalar FF phase
$\delta_\Gamma(t)$ is given just by the ratio of the imaginary
part to the real part of the new numerator as follows
\begin{equation}
   tan \:\delta_\Gamma(t) = \frac{Im [\prod_{i=1}^N (q-q_i)^*\sum_{n=1}^M a_n q^n]}
   {Re [\prod_{i=1}^N (q-q_i)^*\sum_{n=1}^M a_n q^n]}\label{tanphaze}.
\end{equation}

   Further, by using the identity $\delta_\Gamma$ = $\delta_0^0$
following from (\ref{elasticunit2}) and the threshold behavior of
$\delta_0^0$, the following parametrization
\begin{equation}
   tan \:\delta_0^0(t) =
   \frac{A_1q+A_3q^3+A_5q^5+A_7q^7+...}{1+A_2q^2+A_4q^4+A_6q^6+...}\label{newtanphaze}
\end{equation}
or equivalent relation
\begin{equation}
   \delta_0^0(t) =\frac{1}{2i}ln \frac{(1+A_2q^2+A_4q^4+A_6q^6+..)+i(A_1q+A_3q^3+A_5q^5+A_7q^7+...)}
   {(1+A_2q^2+A_4q^4+A_6q^6+..)-i(A_1q+A_3q^3+A_5q^5+A_7q^7+...)}\label{sphase1}
\end{equation}
is obtained from (\ref{tanphaze}), where $A_i$ are all real new
coefficients. The parameter $A_1$ is exactly equal to the $S$-wave
iso-scalar $\pi\pi$ scattering length $a_0^0$.

   One can see directly from (\ref{newtanphaze}) that if the
degree of the numerator is higher than the degree of its
denominator then
\begin{equation}
   \lim _{q \to \infty} \delta_0^0(t) = \frac{\pi}{2}\label{finitelim}.
\end{equation}

   However, if the degree of the numerator in (\ref{newtanphaze})
is lower than the degree of its denominator then
\begin{equation}
   \lim _{q \to \infty} \delta_0^0(t) = 0.
\end{equation}

   The above-mentioned asymptotic behaviors can not be solved beforehand and only a
comparison of (\ref{newtanphaze}) with data on $\delta_0^0(t)$ can
decide what type of pion scalar FF phase representations derived
from either the dispersion relation with one subtraction, or from
the dispersion relation without subtractions, will be the most
suitable in our further considerations.

\section{Analysis of $S$-wave iso-scalar $\pi \pi$ scattering phase shift data}

   There is longdated discussion, what data on $\delta_0^0(t)$ are
more correct, the "down" solution, or the "up" solution.
Fortunately, the latter concerns only in data above $1 {\rm GeV}^2$.

\begin{figure}[h,b]
\includegraphics[scale=.4]{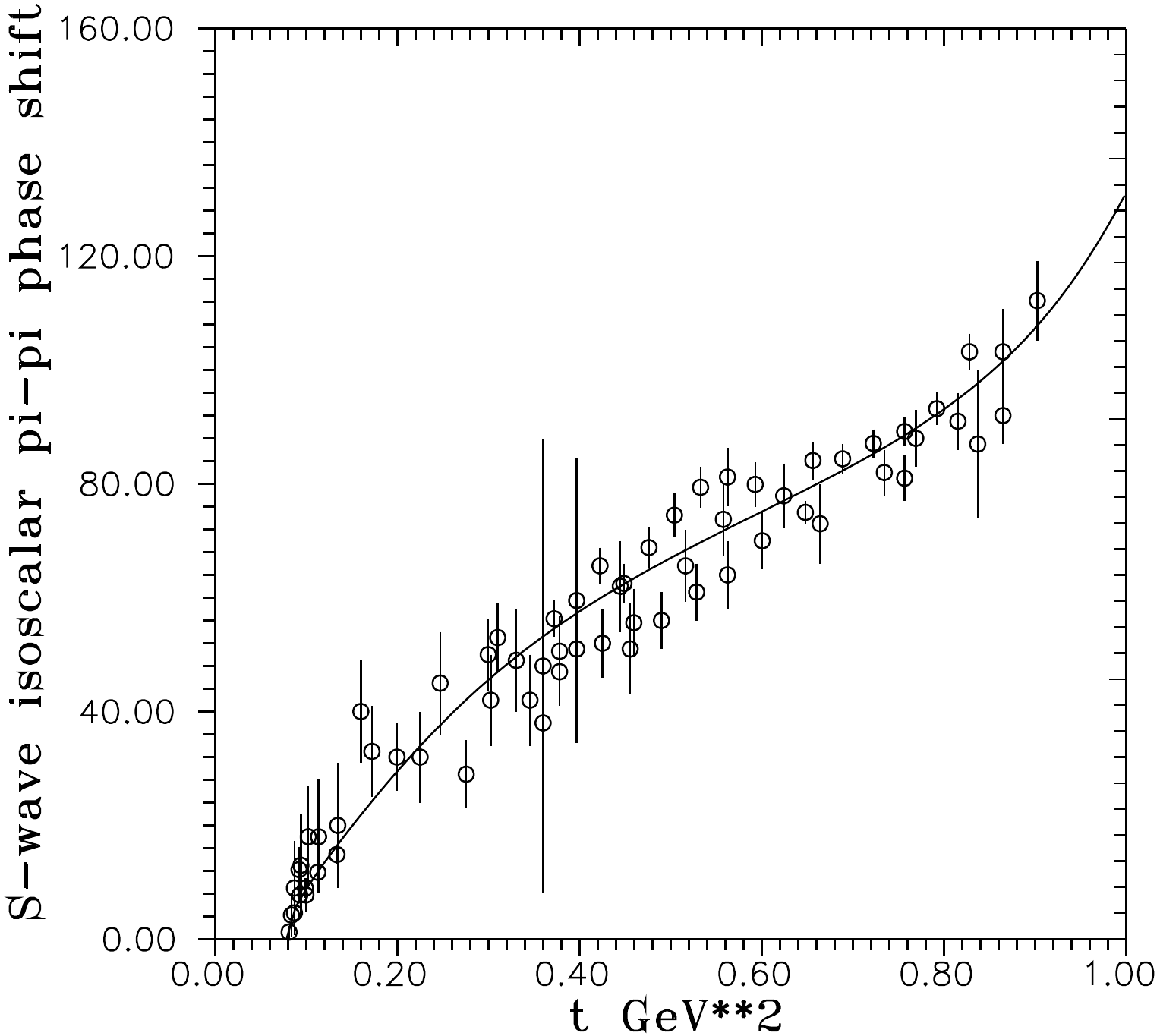}
\caption{\label{Fig.1}Description of the S-wave iso-scalar $\pi\pi$ phase shift
by the [5/4] Pad`e type approximation with the values of
parameters (\ref{numpar})}.
\end{figure}

   As we are interested only for scalar meson resonances below $1
{\rm GeV}^2$, we have collected unambiguous reliable 66 experimental
points (see Fig.\ref{Fig.1}) from \cite{hyams},\cite{estmart},\cite{protop}
and \cite{gunter} at the elastic region without any mutual
discrimination and trying to find the best description of them by
the
\begin{eqnarray}
\delta_0^0(t)=arctan\frac{A_1q+A_3q^3+A_5q^5+A_7q^7+...}
{1+A_2q^2+A_4q^4+A_6q^6+...}\label{sphase2}
\end{eqnarray}
parametrization to be equivalent to (\ref{sphase1}).

   We have carried out the analysis of the data on $\delta_0^0(t)$
successively, starting with the lowest nonzero coefficient $A_1$
and then repeating optimal description of the data always adding
next coefficient to be different from zero. As a result we analyse
the data with one, two, three, etc. parameter expression
(\ref{sphase2}) up to the moment when the minimum of $\chi^2/ndf$
is achieved.

The results are summarized in the following Table

   Number of $A_i$ \qquad  $\chi^2/ndf$

    \qquad 1 \qquad \qquad \qquad 17.75

    \qquad 2 \qquad \qquad \qquad  1.66

    \qquad 3 \qquad \qquad \qquad  1.60

    \qquad 4 \qquad \qquad \qquad  1.49

    \qquad 5 \qquad \qquad \qquad  1.41

    \qquad 6 \qquad \qquad \qquad  1.44

    \qquad 7 \qquad \qquad \qquad  1.50\\
from where one can see immediately that the minimum of
$\chi^2/ndf$ is achieved with 5 coefficients in (\ref{sphase2}).
They are acquirining the following numerical values
\begin{eqnarray}
  \nonumber
  A_1=0.25684 \pm 0.0107; A_3=0.14547 \pm 0.01620;A_5=-.01217 \pm
  0.00070\\\label{numpar}
  A_2=0.02274 \pm 0.02830;A_4=-.01537 \pm
  0.00480\\\nonumber
\end{eqnarray}
and the description of the data by these coefficients is presented
in Fig.\ref{Fig.1} by full line.

   This result (see (\ref{finitelim})) is enough to conclude that we have
to start construction of the pion scalar FF by the dispersion
relation with one subtraction
\begin{eqnarray}
  \Gamma_\pi(t)=1+\frac{t}{\pi}\int^\infty_{4m^2_\pi}\frac{Im
  \Gamma_\pi(t')}{t'(t'-t)}dt'\label{drel1subtr}.
\end{eqnarray}

\section{The phase representation and explicit form of the pion scalar form factor}

   Now, substituting the pion scalar FF elastic unitarity
condition (\ref{elasticunit2}) into the dispersion relation with
one subtraction (\ref{drel1subtr}) one obtains the so-called
Muskelishvili-Omnes integral equation,
\begin{eqnarray}
  \Gamma_\pi(t)=1+\frac{t}{\pi}\int^\infty_{4m^2_\pi}\frac{\Gamma_{\pi}(t')e^{-i\delta_0^0}sin \delta_0^0
   }{t'(t'-t)}dt'\label{muskomn},
\end{eqnarray}
the solution of which is the pion scalar FF phase representation
with one subtraction
\begin{eqnarray}
  \Gamma_\pi(t)=P_n(t)exp[\frac{t}{\pi}\int^\infty_{4m^2_\pi}\frac{\delta^0_0(t')}{t'(t'-t)}dt']\label{phaserepr},
\end{eqnarray}
where $P_n(t)$ is an arbitrary polynomial to be restricted with
$P_n(0)=1$ and its degree must not be higher than
$\delta^0_0(\infty)/\pi$.

   The substitution of $\delta^0_0(t)$ (\ref{sphase2})
in the equivalent form (\ref{sphase1}) with $5$ nonzero
above-mentioned coefficients into the pion scalar FF phase
representation (\ref{phaserepr}) leads to the expression
\begin{eqnarray}
\Gamma_\pi(t)=P_n(t)exp{\frac{(q^2+1)}{\pi
i}\int^\infty_4\frac{q'ln\frac{(1+A_2q'^2+A_4q'^4)+i(A_1q'+A_3q'^3+A_5q'^5)}
{(1+A_2q'^2+A_4q'^4)-i(A_1q'+A_3q'^3+A_5q'^5)}}{(q'^2+1)(q'^2-q^2)}dq'},
\end{eqnarray}
in which  $m_\pi=1$ is assumed. Taking into account the fact that
the integrand is even function of its argument, i.e. it is
invariant under the transformation $q' \to -q'$, the latter
expression can be transformed into the following form
\begin{eqnarray}
\Gamma_\pi(t)=P_n(t)exp{\frac{(q^2+1)}{2\pi
i}\int^\infty_{-\infty}\frac{q'ln\frac{(1+A_2q'^2+A_4q'^4)+i(A_1q'+A_3q'^3+A_5q'^5)}
{(1+A_2q'^2+A_4q'^4)-i(A_1q'+A_3q'^3+A_5q'^5)}}{(q'^2+1)(q'^2-q^2)}dq'}\label{explsmff},
\end{eqnarray}
where the integral is already suitable to be calculated by means
of the theory of residua.

   In order to carry out this program one has to identify all poles of the integrand
and simultaneously calculate complex roots of the polynomial in
the numerator and complex conjugate roots in the denominator under
the logarithm, which generate branch points in $q$-plane.

   Considering the case $q^2<0$ i.e. $q=i\sqrt{\frac{4-t}{4}}\equiv
ib$ one finds the poles of the integrand in $q' = \pm i$ and $q' =
\pm ib$.

   What concerns of the roots of polynomials under the logarithm,
it is clear that it is enough to investigate the roots of the
numerator as the roots of the denominator are complex conjugate to
the roots of the numerator.

   So, let us start with an investigation of the numerator
$(1+A_2q'^2+A_4q'^4)+i(A_1q'+A_3q'^3+A_5q'^5)=0$.

In order to have equation with real coefficients one substitutes $q'=ix$.

Then
  $1-A_1x-A_2x^2+A_3x^3+A_4x^4-A_5x^5=0$

or
  $-x^5+\frac{A_4}{A_5}x^4+\frac{A_3}{A_5}x^3-\frac{A_2}{A_5}x^2-\frac{A_1}{A_5}x+\frac{1}{A_5}=0$

Solutions of the latter equation are the following
\begin{eqnarray*}
  x_1&=&-1.8633297'\\
  x_2&=&0.2832535 -i 3.5830748,\\
  x_3&=&1.2800184 -i 1.3328447,\\
  x_4&=&0.2832535 +i 3.5830748,\\
  x_5&=&1.2800184 +i 1.3328447,
\end{eqnarray*}
from where one finds roots of the numerator and the denominator
under the logarithm of integrand $\phi(q',q)$ to be
\begin{eqnarray}
  \nonumber
  q_1&=&-i1.8633297,\\\nonumber
  q_2&=&-3.5830748+i0.2832535,\\
  q_3&=&-1.3328447+i1.2800184,\\\nonumber
  q_4&=&3.5830748+i0.2832535,\\\nonumber
  q_5&=&1.3328447+i1.2800184,\\\nonumber
\end{eqnarray}
and
\begin{eqnarray}
  \nonumber
  q^*_1&=&-q_1,\\\nonumber
  q^*_2&=&-q_4,\\\label{symroots}
  q^*_3&=&-q_5,\\\nonumber
  q^*_4&=&-q_2,\\\nonumber
  q^*_5&=&-q_3,\\\nonumber
\end{eqnarray}
respectively.

\begin{figure}[t,h,b]
\includegraphics[scale=.4]{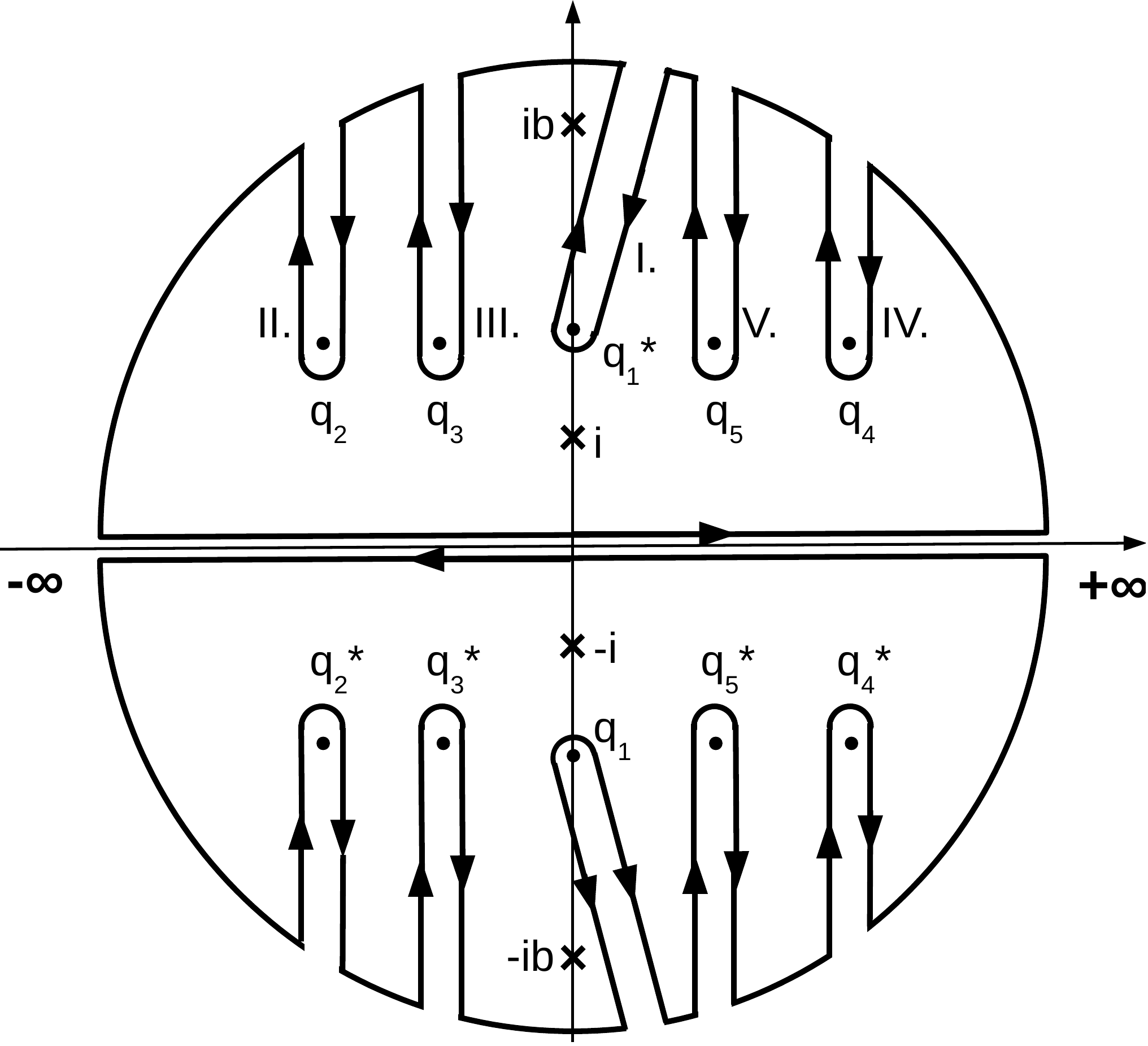}
\caption{\label{Fig.2}Poles ($\times$) and branch points ($\bullet$) of the
integrands $\phi_1(q',q)$ and $\phi_2(q',q)$ with contours of
integrations in the upper and the lower half-planes,
respectively.}
\end{figure}

   Then the integral in (\ref{explsmff}) takes the form
\begin{eqnarray}
  I=\int^\infty_{-\infty}\frac{q'ln\frac{(q'-q_1)(q'-q_2)(q'-q_3)(q'-q_4)(q'-q_5)}
  {(q'-q^*_1)(q'-q^*_2)(q'-q^*_3)(q'-q^*_4)(q'-q^*_5)}}{(q'+i)(q'-i)(q'+ib)(q'-ib)}dq'\label{totint},
\end{eqnarray}
with all singularities of its integrand to be explicitly presented
in Fig.\ref{Fig.2}.

   For an explicit calculation of the latter integral (\ref{totint}) it is convenient
to split it into sum of two integrals
\begin{eqnarray*}
  I = \int^\infty_{-\infty}\frac{q'ln\frac{(q'-q_2)(q'-q_3)(q'-q_4)(q'-q_5)}
  {(q'-q^*_1)}}{(q'+i)(q'-i)(q'+ib)(q'-ib)}dq'+
\end{eqnarray*}
\begin{eqnarray*}
  +\int^\infty_{-\infty}\frac{q'ln\frac{(q'-q_1)}
  {(q'-q^*_2)(q'-q^*_3)(q'-q^*_4)(q'-q^*_5)}}{(q'+i)(q'-i)(q'+ib)(q'-ib)}dq'=I_1+I_2
\end{eqnarray*}
according to singularities to be placed in the upper or lower
half-plane, respectively.

   Let us start to calculate the first integral $I_1$ by the theory of
residua
\begin{eqnarray}
  \oint\frac{q'ln\frac{(q'-q_2)(q'-q_3)(q'-q_4)(q'-q_5)}
  {(q'-q^*_1)}}{(q'+i)(q'-i)(q'+ib)(q'-ib)}dq'=2\pi i \sum_{n=1}^2 Res_n
\end{eqnarray}
where the contour of integration is closed in the upper half-plane
(see Fig.\ref{Fig.2}).

   As the integral on the half-circle is $0$ then
\begin{eqnarray}
 I_1=\int^\infty_{-\infty}\phi_1(q')dq'=2\pi i
 \sum_{n=1}^2 Res_n -
 [-\int_{1^*}+\int_2+\int_3+\int_4+\int_5]\label{expl1in}
\end{eqnarray}
where the integrals on the right-hand side represent contributions
of the cuts generated by the branch points $q_1^*, q_2, q_3, q_4,
q_5$ in Fig.\ref{Fig.2}.

   The residua at the poles $q'=i, q'=ib$ are straightforward to
calculate and they are

\begin{eqnarray}
 Res \phi_1(i,q)= -\frac{1}{2(q^2+1)}ln
 \frac{(i-q_2)(i-q_3)(i-q_4)(i-q_5)}{(i-q_1^*)},
\end{eqnarray}
\begin{eqnarray}
 Res \phi_1(ib,q)= \frac{1}{2(q^2+1)}ln
 \frac{(q-q_2)(q-q_3)(q-q_4)(q-q_5)}{(q-q_1^*)},
\end{eqnarray}
as $ib=q$.

  Now the contributions of the cuts. Let us start with the
contribution of the cut to be generated by the branch point
$q_1^*$.
\begin{eqnarray}
  \int_{1^*} =
  \int_\infty^{q^*_1}\frac{q'ln_+(q'-q_1^*)}{(q'^2+1)(q'^2+b^2)}dq'+
  \int^\infty_{q^*_1}\frac{q'ln_-(q'-q_1^*)}{(q'^2+1)(q'^2+b^2)}dq'=\\\nonumber
  =\int_{q^*_1}\frac{q'}{(q'^2+1)(q'^2+b^2)}[ln_-(q'-q^*_1)-ln_+(q'-q^*_1)]dq'=\\\nonumber
  =-2\pi i\int_{q^*_1}\frac{q'}{(q'^2+1)(q'^2+b^2)}dq'=\\\nonumber
  =-\frac{\pi i}{(b^2-1)}ln\frac{(q^{*2}_1+b^2)}{(q^{*2}_1+1)}\equiv
  \frac{1}{2}\frac{2\pi
  i}{(q^2+1)}ln\frac{(q^{*2}_1-q^2)}{(q^{*2}_1+1)}.
\end{eqnarray}

   Similarly
\begin{eqnarray}
  \int_{j} = -\frac{\pi i}{(b^2-1)}ln\frac{(q^2_j+b^2)}{(q^2_j+1)}\equiv
  \frac{1}{2}\frac{2\pi
  i}{(q^2+1)}ln\frac{(q^2_j-q^2)}{(q^2_j+1)}; \qquad
  j=2,3,4,5.
\end{eqnarray}

   Then the sum of all these partial results according to
(\ref{expl1in}) gives the final result for $I_1$ in the form

\begin{eqnarray}
   I_1 = \frac{1}{2}\frac{2\pi i}{(q^2+1)}ln\frac{(q+q^*_1)}{(q+q_2)(q+q_3)(q+q_4)(q+q_5)}
   \frac{(i+q_2)(i+q_3)(i+q_4)(i+q_5)}{(i+q^*_1)}.\label{finresin1}
\end{eqnarray}

   Similarly one can calculate also the second integral $I_2$ by
means of the theory of residua
\begin{eqnarray}
  \oint\frac{q'ln\frac{(q'-q_1)}{(q'-q^*_2)(q'-q^*_3)(q'-q^*_4)(q'-q^*_5)}
  }{(q'+i)(q'-i)(q'+ib)(q'-ib)}dq'=2\pi i \sum_{n=1}^2 Res_n
\end{eqnarray}
where the contour of integration is closed in the lower half-plane
(see Fig.\ref{Fig.2}).

   As the integral on the half-circle is $0$ then
\begin{eqnarray}
 I_2=\int^\infty_{-\infty}\phi_2(q')dq'=-2\pi i
 \sum_{n=1}^2 Res_n +
 [+\int_{1}-\int_{2^*}-\int_{3^*}-\int_{4^*}-\int_{5^*}]\label{expl2in}.
\end{eqnarray}

    The residua at the poles $q'=-i, q'=-ib$ take the form
\begin{eqnarray}
 Res \phi_2(-i,q)= -\frac{1}{2(q^2+1)}ln
 \frac{(-i-q_1)}{(-i-q^*_2)(-i-q^*_3)(-i-q^*_4)(-i-q^*_5)},
\end{eqnarray}
\begin{eqnarray}
 Res \phi_2(-ib,q)= \frac{1}{2(q^2+1)}ln
 \frac{(-q-q_1)}{(-q-q^*_2)(-q-q^*_3)(-q-q^*_4)(-q-q^*_5)},
\end{eqnarray}
as $ib=q$.

   The contribution of the cut to be generated by the branch point
$q_1$ is
\begin{eqnarray}
  \int_{1} =
  \int_\infty^{q_1}\frac{q'ln_+(q'-q_1)}{(q'^2+1)(q'^2+b^2)}dq'+
  \int^\infty_{q_1}\frac{q'ln_-(q'-q_1)}{(q'^2+1)(q'^2+b^2)}dq'=\\\nonumber
  =\int_{q_1}^\infty\frac{q'}{(q'^2+1)(q'^2+b^2)}[ln_-(q'-q_1)-ln_+(q'-q_1)]dq'=\\\nonumber
  =-2\pi i\int_{q_1}^\infty\frac{q'}{(q'^2+1)(q'^2+b^2)}dq'=\\\nonumber
  =-\frac{\pi i}{(b^2-1)}ln\frac{(q^2_1+b^2)}{(q^2_1+1)}\equiv
  \frac{1}{2}\frac{2\pi
  i}{(q^2+1)}ln\frac{(q^2_1-q^2)}{(q^2_1+1)}.
\end{eqnarray}

   Similarly
\begin{eqnarray}
  \int_{j^*} = -\frac{\pi i}{(b^2-1)}ln\frac{(q^{*2}_j+b^2)}{(q^{*2}_j+1)}\equiv
  \frac{1}{2}\frac{2\pi i}{(q^2+1)}ln\frac{(q^{*2}_{j}-q^2)}{(q^{*2}_{j}+1)}; \qquad
  j=2,3,4,5.
\end{eqnarray}
\begin{figure}[t,h,b]
\includegraphics[scale=.4]{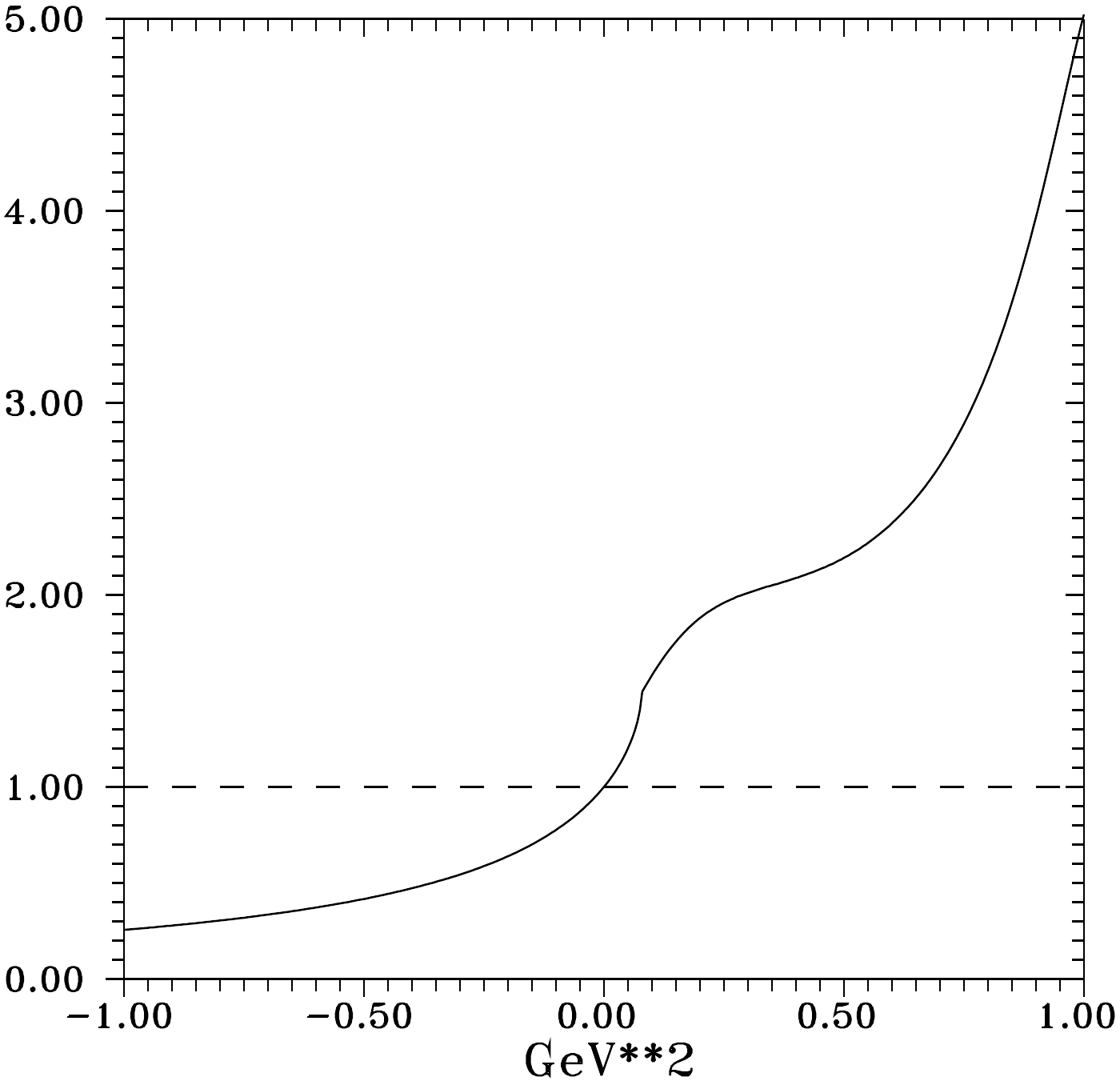}
\caption{\label{Fig.3}Behavior of the pion scalar form factor in the region $-1
{\rm GeV}^2 < t < 1 {\rm GeV}^2$ }
\end{figure}
   Then the sum of all these partial results according to
(\ref{expl2in}) gives the comprehensive result for $I_2$ in the
form
\begin{eqnarray}
  I_2=\frac{1}{2}\frac{2\pi i}{(q^2+1)}\nonumber
  [ln\frac{(q+q_1)}{(q+q^*_2)(q+q^*_3)(q+q^*_4)(q+q^*_5)}
  \frac{(i+q_2^*)(i+q_3^*)(i+q_4^*)(i+q_5^*)}{(i+q_1)}+\\
  +ln\frac{q_1^2-q^2}{q_1^2+1}-
  \frac{q_2^{*2}-q^2}{q_2^{*2}+1}-\frac{q_3^{*2}-q^2}{q_3^{*2}+1}-
  \frac{q_4^{*2}-q^2}{q_4^{*2}+1}-
  \frac{q_5^{*2}-q^2}{q_5^{*2}+1}]
\end{eqnarray}
from where by using the relations (\ref{symroots}) finally one
gets
\begin{eqnarray}
   I_2 = \frac{1}{2}\frac{2\pi i}{(q^2+1)}ln\frac{(q+q^*_1)}{(q+q_2)(q+q_3)(q+q_4)(q+q_5)}
   \frac{(i+q_2)(i+q_3)(i+q_4)(i+q_5)}{(i+q^*_1)}\label{finresin2}.
\end{eqnarray}
The sum of (\ref{finresin2}) with (\ref{finresin1}) represents the
total integral

\begin{eqnarray}
  I=\frac{2\pi i}{(q^2 +1)}ln
  \frac{(q-q_1)}{(q+q_2)(q+q_3)(q+q_4)(q+q_5)}\frac{(i+q_2)(i+q_3)(i+q_4)(i+q_5)}{(i-q_1)}.
\end{eqnarray}
If the latter is substituted into the pion scalar FF phase
representation (\ref{explsmff}) one obtains an explicit form for
the pion scalar FF $\Gamma_\pi(t)$
\begin{eqnarray}
   \Gamma_\pi(t)=P_n(t)\frac{(q-q_1)}{(q+q_2)(q+q_3)(q+q_4)(q+q_5)}\frac{(i+q_2)(i+q_3)(i+q_4)(i+q_5)}{(i-q_1)},
\end{eqnarray}
which behavior graphically is presented in Fig.\ref{Fig.3}.

   The  $-q_3$ and $-q_2$ poles of $\Gamma_\pi(t)$ on the second Riemann sheet
in $t$-variable correspond to $f_0(500)$ and $f_0(980)$ scalar
meson resonances, respectively.

  Their masses and widths are determined to be

  ${\bf m_{f_0(500)}=(360\pm 33) {\rm MeV}}$, \quad ${\bf \Gamma_{f_0(500)}=(587\pm 85)
  {\rm MeV}}$,

  ${\bf m_{f_0(980)}=(957\pm 72) {\rm MeV}}$, \quad ${\bf \Gamma_{f_0(980)}=(164\pm 142)
  {\rm MeV}},$\\
  where the errors correspond to the transferred errors of the
  coefficients of (\ref{numpar}).

   The parameters of $f_0(500)$ can be compared with other determinations
presented in the following Table

   $m_\sigma=441 {\rm MeV}$, \quad $\Gamma_\sigma=544 {\rm MeV}$, \cite{capr1}

   $m_\sigma=474 {\rm MeV}$, \quad $\Gamma_\sigma=508 {\rm MeV}$, \cite{garcia}

   $m_\sigma=463 {\rm MeV}$, \quad $\Gamma_\sigma=508 {\rm MeV}$, \cite{capr2}

   $m_\sigma=443 {\rm MeV}$, \quad $\Gamma_\sigma=432 {\rm MeV}$, \cite{oller}

   $m_\sigma=452 {\rm MeV}$, \quad $\Gamma_\sigma=520 {\rm MeV}$, \cite{mennes}

   $m_\sigma=453 {\rm MeV}$, \quad $\Gamma_\sigma=542 {\rm MeV}$, \cite{palaez}

   $m_\sigma=457 {\rm MeV}$, \quad $\Gamma_\sigma=558 {\rm MeV}$, \cite{gar-mar}.

As one can see immediately from this Table, its mass in these
determinations is slightly higher, whereas the width is lower,
than in our model independent approach.

   Finally one can only say, that if more precise data on the
S-wave iso-scalar $\pi\pi$ scattering phase shift is available,
more precise parameters of $f_0(500)$ and $f_0(980)$ can be found
in our model independent method of their determination.

\section{CONCLUSIONS}

   The unitary and analytic approach has been applied for a prediction
of the pion scalar FF behavior in elastic region, in the framework
of which only the experimental data on S-wave iso-scalar $\pi\pi$
scattering phase shift in elastic region were used to determine
the $f_0(500)$ and $f_0(980)$ scalar meson parameters in a model
independent way.

   The support of the Slovak Grant Agency for Sciences VEGA under
Grant No. 2/0158/13 and of the Slovak Research and Development
Agency under the contract No. APVV-0463-12 is acknowledged.

\end{document}